# Superconductivity in the Intercalated Graphite Compounds $C_6$Yb and $C_6$Ca

Thomas E Weller‡, Mark Ellerby‡, Siddharth S Saxena†, Robert P Smith† and Neal T Skipper‡

*‡ Department of Physics and Astronomy, University College of London, Gower Street, London WC1E 6BT, UK*

*† Cavendish Laboratory, University of Cambridge, Madingley Road, Cambridge CB3 0HE, UK*

**In this letter we report the discovery of superconductivity in the isostructural graphite intercalation compounds $C_6$Yb and $C_6$Ca, with transition temperatures of 6.5K and 11.5K respectively. A structural characterisation of these compounds shows them to be hexagonal layered systems in the same class as other graphite intercalates. If we assume that all the outer *s*-electrons are transferred from the intercalant to the graphite sheets, then the charge transfer in these compounds is comparable to other superconducting graphite intercalants such as $C_8$K[1,2]. However, the superconducting transition temperatures of $C_6$Yb and $C_6$Ca are up to two orders of magnitude greater. Interestingly, superconducting upper critical field studies and resistivity measurements suggest that these compounds are significantly more isotropic than pure graphite. This is unexpected as the effect of introducing the intercalant is to move the graphite layer further apart.**

Graphite is a form of carbon in which the atoms are arranged hexagonally in two-dimensional sheets. Pure graphite is a very interesting compound displaying unusual properties; for example, the two-dimensional sheets form a semi-conductor in which the band gap is zero. In bulk graphite these two dimensional carbon layers are held together via weak van der Waals forces. It is this feature which makes it possible to introduce metal atoms in between the layers of carbon. The process of introducing these metal atoms is known as intercalation and often results in the formation of graphite intercalation compounds in which an ordered structure is formed. These graphite intercalation compounds provide an excellent laboratory in which to study low-dimensional electronic systems in a controlled fashion. In particular, the introduction of these metal atoms is thought both to donate electrons to carbon layers and to change the spacing of these layers. These processes result in a marked modification of both the physical and electronic properties[3-5]. In particular, unlike pure graphite, some graphite intercalation compounds are found to superconduct. The first of these to be reported [1,2] was $C_8K$, which has a superconducting transition temperature of 0.15K. Interestingly, while the metastable high pressure phase $C_2Li$ exhibits a superconducting transition at 1.9K [6], the compounds $C_6Li$ and $C_3Li$ are found[5] to not superconduct down to the lowest measured temperatures. In all these compounds the transfer of charge from the metal to the graphite is thought to play an important role in the superconductivity. However, we see that there must be additional factors at work as both potassium (K) and lithium (Li) would be expected to donate one electron each to the graphite and $C_8K$ superconducts while $C_6Li$ does not. This non-trivial behaviour showed that the fabrication and study of different graphite intercalation compounds would be worthwhile. Therefore, we have fabricated the isostructural intercalation compounds $C_6Yb$ and $C_6Ca$. Here we present results demonstrating the existence of superconductivity in these compounds together with a structural determination (see method and figure 1) showing the formation of ordered structures. The structure is a hexagonal layered structure (*P6₃/mmc*) in which the intercalant atoms form a triangular array between every graphite layer (stage 1 intercalation). The alternate carbon and metal layers have an AαAβ registration[7] where the A represents the carbon layers and the α and β the intercalant layers. The superconducting transition temperatures ($T_{sc}$) for these compounds are 6.5K in $C_6Yb$ and 11.5K in $C_6Ca$.

Two of the principle signatures of superconductivity are the absence of electrical resistivity and the development of a diamagnetic moment below the ordering transition. Figure 2 shows the results of resistivity and DC magnetisation measurements made on samples of $C_6Yb$ and DC magnetisation measurements on $C_6Ca$. The results for $C_6Yb$ (figures 2a and 2b) show a clear transition at 6.5K in the magnetization and the resistivity, both of which support the existence of superconductivity. The transition is well defined, having width of 0.2K in the resistivity. These intercalation compounds are very difficult to make (see methods), however we have managed to make samples in which X-ray studies show that over 13% of the final volume of the sample is $C_6Yb$. However, field-dependent magnetization measurements made parallel to **c\***-axis of $C_6Yb$ imply that the superconducting volume fraction is approximately 90%. This difference was resolved by scanning electron microscopy (SEM) studies which revealed that the intercalation process creates a "shell" of the intercalant with a core of pristine graphite. It is important to stress that the subsequent cleaving of several layers up to 300 μm from these samples did not remove superconductivity. The magnetization measurement shown in figure 2b, with the field of 50Oe applied parallel to the **c\***-axis was performed on a disk shaped sample. The zero field cooled (ZFC) data reveals the flux expulsion and subsequent flux threading as the temperature is increased. The field cooled (FC) measurements, when compared with the ZFC result, exhibit only partial flux expulsion. In fields exceeding the superconducting upper critical field ($H_{C2}$) we find a weak paramagnetic signal. The origins of this paramagnetic moment are difficult to attribute, but we do have X-ray evidence (figure 1) showing a contamination of less than 1% $Yb_2O_3$ which is known to have an ordered moment below 2.4K[8].

The $C_6Ca$ magnetisation and resistivity results are shown in figure 2c. In the magnetization we see a clear diamagnetic onset at 11.5K in a field of 50Oe, but with no saturation of diamagnetism down to 2K. In addition, the preliminary resistivity measurements demonstrate that the resistivity goes to zero below the transition temperature. However, extreme air sensitivity and difficulties in preparation of this compound [11] have prevented detailed transport and magnetisation measurements thus far. The fact that the magnitude of the diamagnetic moment is about 100 times smaller than in $C_6Yb$ also points to reduced sample quality. We are able to conclude that $C_6Ca$ superconducts below 11.5K.

Figure 2d presents the magnetic phase diagram for $C_6Yb$ inferred from magnetization measurements made with applied magnetic field in the plane of the layers (ab-plane) and perpendicular to them (**c***-axis). The lower critical field $H_{C1}$ is approximately the same for both geometries, whilst the upper critical field $H_{C2}$ is clearly anisotropic. The anisotropy parameter of $\Gamma_{H_{C2}}$ given by $H_{H_{C2}}(\perp_c)/H_{H_{C2}}(\|_c)$ is approximately 2 across the temperature range below $T_{SC}$. In the Ginzburg-Landau theory, this anisotropy depends solely on the ratio of the electron masses along the two symmetry directions. Calculation of the ratio of the effective masses for pure graphite yields a value for $\Gamma_{H_{C2}}$ of 7. A comparison of these two values implies the Fermi surface is more three-dimensional in $C_6Yb$ compared to pure graphite. This observation is consistent with normal state resistivity measurements above $T_{SC}$. These measurements show a distinct anisotropy depending if the current is applied in the **c***-axis or in basal plane. This anisotropy can be quantified by the ratio of the **c***-axis resistivity to the ab-plane resistivity, at room temperature this ratio is 100 for $C_6Yb$. This ratio is smaller than the pure graphite from which $C_6Yb$ was made [10], which at room temperature, has an anisotropy ratio of around $10^4$. Further evidence for this change in Fermi surface comes from the temperature-dependent resistivity parallel to the **c***-axis. This reveals a significantly different behaviour from that seen in pure graphite. In pure graphite the resistivity is observed[11] to increase with decreasing temperature reaching a maximum at ∼ 50 K, whilst $\rho_{//c}(T)$ in $C_6Yb$ is found to decrease from room temperature to the transition temperature. These three observations taken together lead us to believe that the Fermi surface in $C_6Yb$ is more isotropic than that found in pure graphite.

Our results leave us with a significant question. *Is a simple charge transfer model, utilised to understand earlier studies[12,13], adequate in explaining the results reported here?* This question arises from the following observations. If charge transfer from the metal to the carbon atoms was the most important affect leading to superconductivity in these systems then we would expect $C_3Li$ to become superconducting, since the charge transfer is comparable to that of $C_6Yb$ and $C_6Ca$, with 1/3 of an electron per carbon being transferred. In fact [5] $C_3Li$ is not superconducting. Also, in the case of $C_2Li$ [6], where the charge transfer is greater (1/2 e per carbon), the transition temperature is still a factor of 3 and 6 smaller than in $C_6Yb$ and $C_6Ca$, respectively. The case [14] of $C_2Na$, which is superconducting at 5K, is less enlightening as the structure and thus the physics of the compound has so far been intractable. In addition, in a conventional phonon mechanism we would not expect an order of magnitude change in $T_{SC}$ on going from a charge transfer of 1/8 of an electron per carbon to 1/3 of an electron per carbon. Therefore, our results have highlighted that there is no clear trend between the amount of charge transferred and the superconducting transition temperature. Thus, there is a demand for a renewed theoretical effort to place our findings in context with previous experimental results on the superconducting graphite intercalation compounds.

In summary, we have synthesised two new isostructural graphite-based superconductors, $C_6Yb$ and $C_6Ca$, with superconducting transition temperatures of 6.5K and 11.5K, respectively. These are unprecedented in the field of graphite intercalates. In addition, we have evidence to suggest that these compounds are more isotropic than pure graphite. This trend to become more isotropic is contrary to a simple picture in which introducing the metal atoms increases the spacing between the graphite layers and so would be expected to make graphite intercalants more, rather than less, two dimensional. In trying to understand this problem further, we will be able to exploit the weak van der Waals bonding between the graphene sheets to explore the impact of interlayer coupling using "tuning" parameters such as hydrostatic pressure or doping with different metals. In addition, this work may also be of more general importance in understanding and exploring superconductivity in the quasi-one dimensional system formed in single walled carbon nanotubes [15].

**Method**

The samples were prepared using the well established vapour transport process[5]. Highly oriented pyrolytic graphite, grade ZYA from Advanced Ceramics Corp. (Cleveland, Ohio) was used as the host for the reaction. The ytterbium was supplied by Goodfellow with a purity of 99.9%, whilst the calcium was supplied by Aldrich with a purity of 99.5%. Before the vapour transport process the graphite has a hexagonal layered structure ($P6_3/mmc$) with an AB registration of graphene layers. Air abrasion techniques were used to prepare the edges of the graphite to produce open galleries. The surfaces were then cleaved to provide clean graphite in a variety of shapes for different probes, and having a thickness of between 0.1mm and 0.25mm. The graphite and the intercalant (Yb or Ca) were sealed at opposite ends of a quartz tube and then treated using a two zone technique with an appropriate thermal gradient [7,9]. This allows the ytterbium or calcium to enter the graphite galleries.

Initial X-ray diffraction studies revealed the presence of graphite, $Yb_2O_3$, YbO and pure Yb in samples of $C_6Yb$. Corresponding measurements for $C_6Ca$ did not reveal the presence of any oxides or free Ca, but does show the clear presence of graphite as a secondary phase. Subsequent SEM of the $C_6Yb$ samples established that there was surface contamination samples that may be removed through abrasion. Figure 1 shows an example of the X-ray patterns for the $C_6Yb$ samples after abrasion of the surface. The geometry employed is a standard technique[3] for identification of the staging of graphite intercalation compounds, where the stage represents the number of graphene layers between intercalant layers; this technique allows only the observation of *(00l)* reflections as indexed. We have been able to model[7] the intercalation phases present in the X-ray data by considering a stacking registry A$\alpha$A$\beta$ with the structure $C_6M$ shown inset in figure 1.

We acknowledge M. Baxendale and M. Laad for their fruitful comments, suggestions and discussions. In the preliminary development of the sample preparation we are indebted to C. Gallagher, J. Goodyer and S.L. Molodtsov (Dresden TU). We thank K.S. Lee (London Centre for Nanotechnology UCL) and Institute of Archaeology, UCL for SEM-EDAX studies on the final samples. We are also grateful to the Eastman Dental Institute for cutting of graphite samples. We are grateful to G. Aeppli, G. Csanyi, A.H. Harker, A. Kusmartseva, D.F. McMorrow, A. H. Nevidomskyy, C. Pickard, P.B. Littlewood, G.G. Lonzarich, J. Loram, S. Ozcan, Ch. Renner, S. Rowley and B.D. Simons for useful discussions. We would also like to thank the EPSRC, CCLRC (ISIS), The Royal Society, Jesus and Trinity Colleges of the University of Cambridge for financial support.

Correspondence and requests for materials should be addressed to Dr. S.S. Saxena (e-mail: sss21@cam.ac.uk) or Dr. Mark Ellerby (e-mail: mark.ellerby@ucl.ac.uk).

**Figure 1 X-ray diffraction (XRD) pattern of HOPG intercalated with Yb.** This data was taken using a Bragg-Brentano geometry with Cu-K$_\alpha$ radiation. The consequence of this is that only the (00*l* ) peaks may be sampled. Inset

showing the derived structure modelled [7] using a stage-1 GIC with a **c*** -axis sandwich depth (C-Yb-C) of 4.57Å. Measurements for $C_6Ca$ have a sandwich depth (C-Ca-C) of 4.60Å. In this structure the graphite sheets have a A–A registration whilst the ytterbium and calcium have an α–β registration. Calculations based on peak intensities reveal that 13% of the sample volume fraction is made up of $C_6Yb$. From our analysis we find that the contamination of $Yb_2O_3$ is less than 1% after surface abrasion.

**Figure 2 Temperature dependence of the magnetization and electrical resistivity for $C_6Yb$ and $C_6Ca$.** Magnetization measurements for $C_6Yb$ and $C_6Ca$ shown in (a) and (c) respectively. These measurements were made with a 50 Oe field applied parallel to the **c*** -axis. These figures reveals the onset of flux expulsion in both the zero field cooled (ZFC) measurement and the field cooled (FC) measurement. The resistivity measurement for $C_6Yb$ is shown in (b). There is a clear drop to zero resistivity indicating the existence of superconductivity. Figure 2 (d) is the superconducting phase diagram for $C_6Yb$. This diagram is compiled from results of the magnetization study. In both geometries the sample appears to be a type-II superconductor. There is little if any anisotropy in $H_{C1}$ for the two geometries, whilst there is a clear anisotropy in $H_{C2}$.

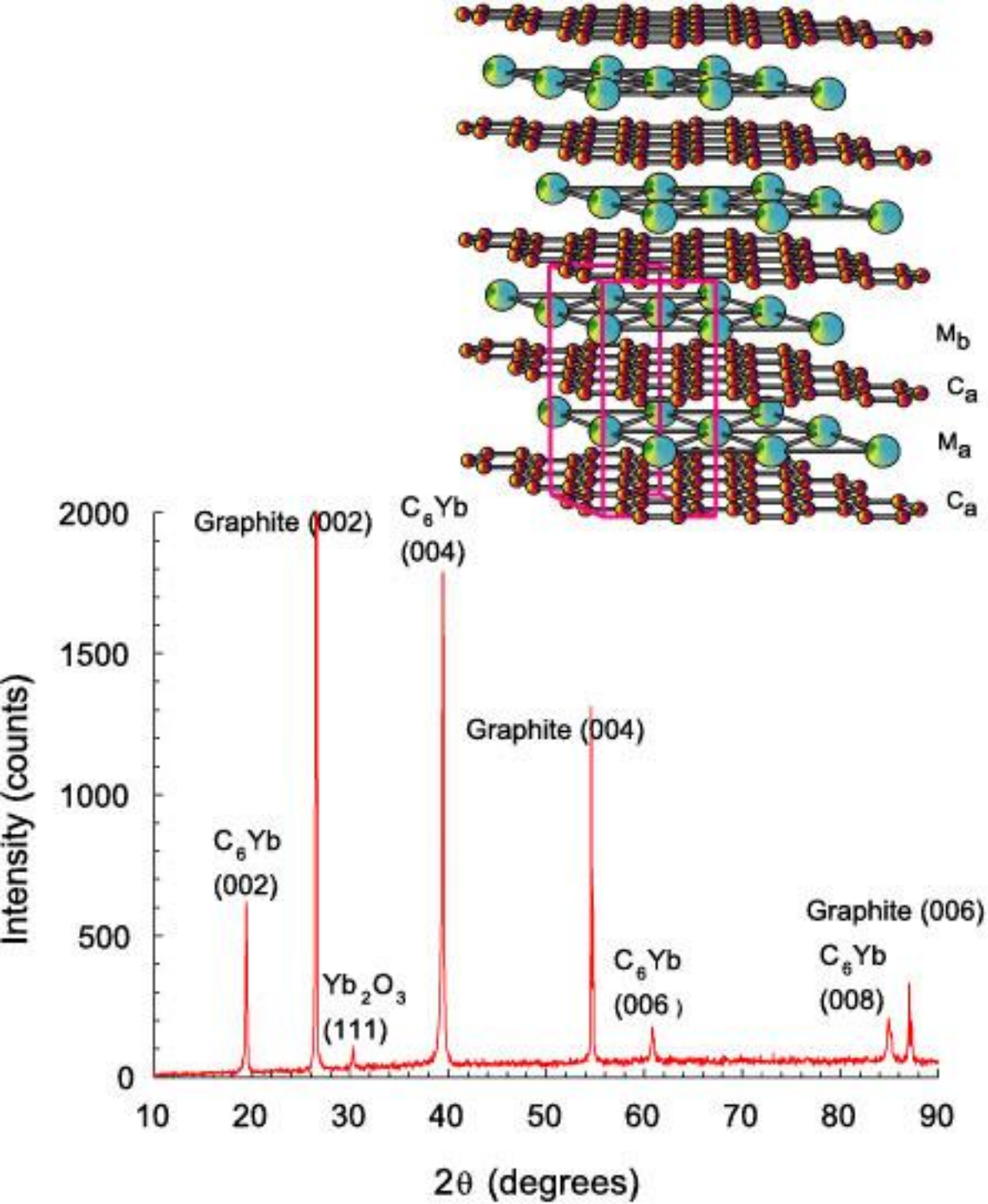

Figure 1

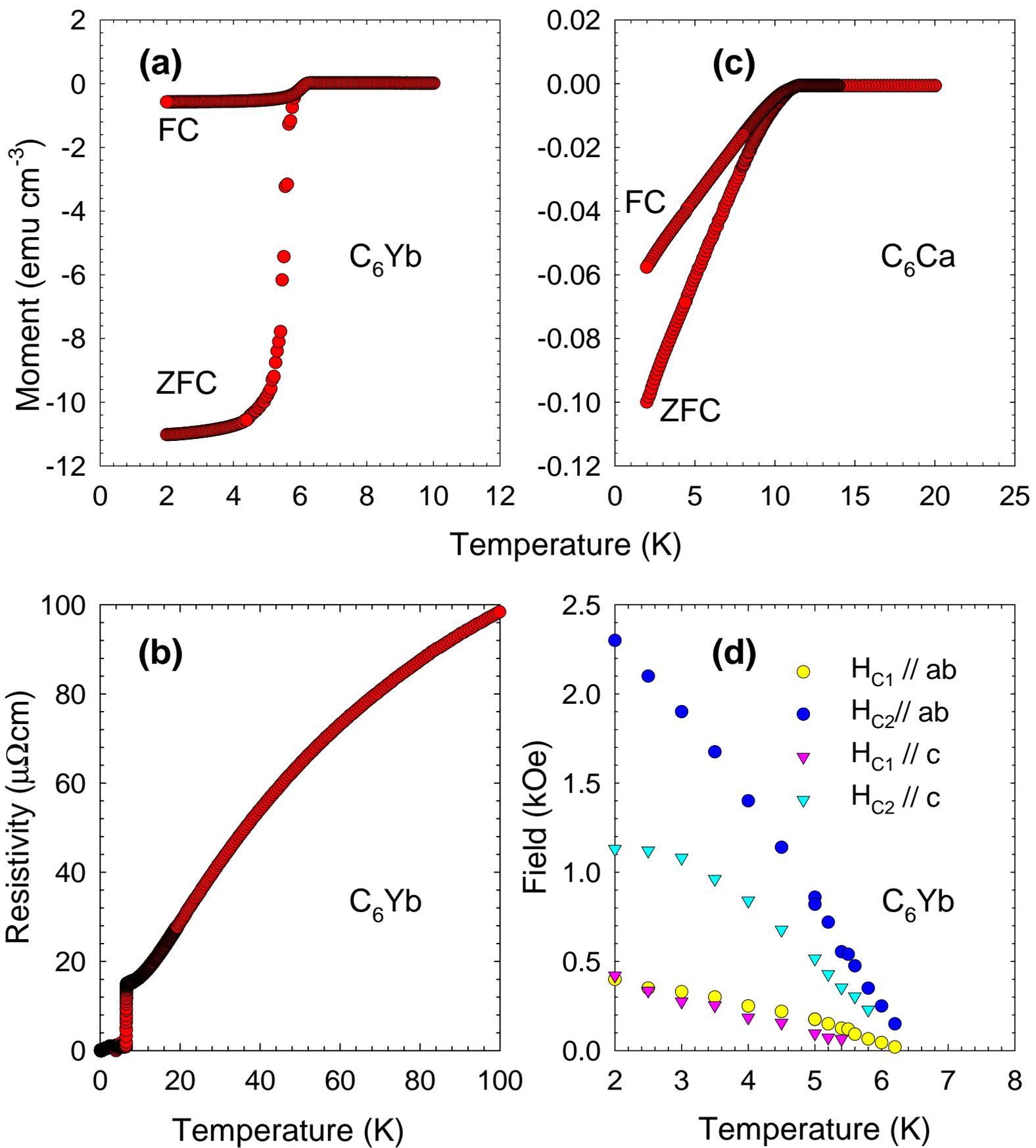

Figure 2